\documentclass[aps,prd,twocolumn,showpacs,floatfix,preprintnumbers,nofootinbib
]{revtex4-1}

\usepackage{graphicx}
\usepackage{hyperref}
\usepackage{color}
\usepackage{natbib}
\usepackage{multirow}
\usepackage{amsmath}
\usepackage{amssymb}
\usepackage{epsfig}
\usepackage{textcomp}
\usepackage{color}
\usepackage{ulem}

\def\theta{\vartheta}

\newcommand{\be}{\begin{equation}}
\newcommand{\ee}{\end{equation}}
\newcommand{\ba}{\begin{eqnarray}}
\newcommand{\ea}{\end{eqnarray}}
\newcommand{\lsim}   {\mathrel{\mathop{\kern 0pt \rlap
  {\raise.2ex\hbox{$<$}}}
  \lower.9ex\hbox{\kern-.190em $\sim$}}}
\newcommand{\gsim}   {\mathrel{\mathop{\kern 0pt \rlap
  {\raise.2ex\hbox{$>$}}}
  \lower.9ex\hbox{\kern-.190em $\sim$}}}

\begin{document}

\title{Comment on  arXiv:1408.0288v2 [hep-ph] and Phys.\ Rev.\ D 90, 085017 (2014), ``A new evaluation of the antiproton production cross section for cosmic ray studies",  Mattia di Mauro {\it et al.} }
\author{R.\ H.\ Bernstein$^{1}$}
\affiliation{$^{1}$Fermi National Accelerator Laboratory, P.O.\ Box 500, Batavia IL 60510 USA}

\begin{abstract}
The referred-to paper on antiproton production contains two errors: an error in the value reported for one of the fit parameters and an error in units.  I give the errors and the corrected values when possible.
\end{abstract}


\maketitle

The authors of Ref.~\cite{DiMauro:2014} have written  an interesting and useful paper on parameterizations of antiproton production.  It  contains two errors, which may be a problem for researchers that attempt to use the published parameterizations.  These parameterizations are relevant for the design of the Mu2e and COMET experiments searching for muon-to-electron conversion, $\mu^-N \rightarrow e^-N$, since antiproton production can lead to a background.\cite{mu2eAndComet} 

Specifically,
\begin{enumerate}
\item The parameter D4 in Table 2 of Ref.~\cite{DiMauro:2014} is quoted as $0.0510(0.050)$. In comparison, Ref.~\cite{Duperray:2003} quotes $0.5121(27)$  for Ref.~\cite{Duperray:2003}'s fit to the same functional form.
 Upon corresponding with an author of Ref.~\cite{DiMauro:2014} they stated the correct value should be $0.51024$, $\times 10$ higher.\cite{DiMauroEmail:2017}
  \item Eqn.~10 has a units mistake.  Eqn.~10 reads:
\begin{eqnarray}
\sigma_{\rm in}(s) &=& \sigma_o \left[ 1 - 0.62 e^{-\frac{E_{\rm inc}(s)}{0.2}} \sin \left( \frac{10.9}{E^{-0.28}_{\rm inc}(s)}\right) \right]
\end{eqnarray}
and the text says ``where $E_{\rm inc}(s)$ is the incident kinetic energy in GeV",  and then 
 refers back to ~\cite{Duperray:2003}.  If we follow \cite{Duperray:2003} we find it refers to \cite{Letaw:1983} and specifically to Eqn.~5 of that paper.  The original equation, with energies in MeV,  read:
\begin{eqnarray}
\sigma_{\rm in}(s) &=& \sigma_o \left[ 1 - 0.62 e^{-\frac{E_{\rm inc}(s)}{200}} \sin \left( \frac{10.9}{E^{-0.28}_{\rm inc}(s)}\right) \right]
\end{eqnarray}
and inspection shows the units have been changed in the exponential but not in the sinusoidal term.  Again, correspondence with an author of Ref.~\cite{DiMauro:2014} confirms the error.\cite{DiMauroEmail:2017}

\end{enumerate}

This manuscript has been authored by Fermi Research Alliance, LLC under Contract No. DE-AC02-07CH11359 with the U.S. Department of Energy, Office of Science, Office of High Energy Physics.


\begin{thebibliography}{99}
\bibitem{DiMauro:2014}{M.\ di Mauro, F.\ Donato, A.\ Goudelis, and P. D.\ Serpico,
Phys.\ Rev.\ D 90, 085017 (2014); also see \url{https://arxiv. org/abs/1408.0288.}}
\bibitem{mu2eAndComet}{The Mu2e Technical Design Reprt can be found at L.\ Bartoszek {\it et al.}, arXiv:1501.05241, Fermilab-TM-2594, Fermilab-DESIGN-2014-01;  the COMET Technical Design Report, at \url{http://comet.kek.jp/Documents_files/IPNS-Review-2014.pdf}.}
\bibitem{Duperray:2003}{R.\ P.\  Duperray, C.-Y.\ Huang, K. V.\ Protasov, and
M. Bu\'{e}nerd, Phys.\ Rev.\ D 68, 094017 (2003); also see \url{arXiv:astro-ph/0305274v1}}.
\bibitem{DiMauroEmail:2017} {R.\  H.\ Bernstein and M.\ di Mauro, emails of  3--4 January 2017.}
\bibitem{Letaw:1983}{J.\ Letaw, R.\ Silberberg, and C. H.\ Tsao, Astropart.\ J.\ Suppl.\ 51, 271 (1983). }
\end{thebibliography}
 \end{document}